\documentclass[12pt,a4paper,twoside,fleqn,epsf]{article}
\usepackage{graphicx}
\def\be{\begin{equation}}
\def\ee{\end{equation}}
\def\bea{\begin{eqnarray}}
\def\eea{\end{eqnarray}}
\def\nnb{\nonumber}

\def\indvert{\vphantom{\frac{\frac{|}{|}}{\frac{|}{|}}}}

\begin{document}
\thispagestyle{empty}

\vspace{0.5cm}
\begin{center}
{\Large Determining the quantum numbers of excited heavy mesons}\\
\vspace{0.5cm}
{G. Eilam and F. Krauss}\\
{Technion--Israel Institute of Technology}\\
{32000 Haifa, Israel}\\


\parbox{10cm}{\centerline{Abstract}
\vspace{0.5cm}{\footnotesize We discuss the decays $X^*\to Xe^+e^-$ 
(``Dalitz decays'') of 
excited heavy mesons into their ground states and an electron--positron pair. 
We argue that the measurement of the invariant mass spectrum of the 
lepton pair 
gives clear indication on the quantum numbers of the excited meson and thus
provides an experimental test of the quark model predictions.}}\\
\vspace{0.5cm}
\end{center}

\noindent We investigate radiative 
decays of excited heavy mesons with charm and beauty,
i.e. $D^{*0}$, $D^{*+}_s$, $B^{*0}$ and $B^{*0}_s$,
into their ground 
state plus an electron--positron pair. There are several reasons that led us to
investigate these decays:

\begin{enumerate}
\item The quantum numbers $J$ and $P$ of the excited mesons given in the 
{\em Review of Particle Physics} \cite{RPP} are either merely predictions of 
the quark model, 
which gives for the particles discussed here $J^P=1^-$, or at best the quantum numbers need 
confirmation \footnote{We would like to mention here, that for the
$B$--meson even the quantum numbers of the ground state are not 
known experimentally; nevertheless we assume $J^P(B)=0^-$.}.
\item The quantum numbers of some even higher heavy meson states, 
depend on the correct assignments for the $X^*$'s which are the focus
of the present work.
\item The central values of the 
branching ratios for $D^{*0}\to D^0\gamma$ and $D^{*0}\to D^0\pi$ 
sum up to exactly 100 $\%$ \cite{RPP}. 
Only closer inspection of both relevant experimental 
papers \cite{exp1} reveals that this was one of the assumptions underlying 
the analyses. A similar assumption was made for $D^{*\pm}_s$, where again it was assumed 
that branching ratios for $D^{*+}_s\to D^+_s\gamma$ and $D^{*+}_s\to D^+_s\pi^0$ 
sum to 100 $\%$ \cite{exp2}. 
Although the 
current errors on the individual branching ratios are clearly higher than the branching ratios 
for the decay into a Dalitz pair which, as we will see below, 
are of the order of $
0.5\%$ of their real photon emission counterparts, 
in the long term this channel cannot be neglected. 
Especially in light of future precision measurements in 
the $B$--sector, a precise
tracking of all outgoing $D$'s is required. 
In other words, it is important
to know which decay of $X^*$ has the closest branching ratio to the decays
that have been already measured. 
Let us remark, that for the $B^*$
decay, for which the dominant mode is $B^* \to B \gamma$, its
decay into $B e^+ e^-$ presented here has a much larger rate 
(about $0.5\%$, see below)
than the decay $B^* \to B \gamma \gamma$, recently discussed in \cite{Paul}.
\item The only recent theoretical analysis of an $X^*\to Xe^+e^-$ decay 
was performed for $D^*\to De^+e^-$
\cite{Aliev} and yielded $R \approx 0.001$, where the ratio $R$ is defined as
\bea
R=\frac{\Gamma(X^*\to Xe^+e^-)}{\Gamma(X^*\to X\gamma)} \label{ratio}
   =\frac{\Gamma_{ee}}{\Gamma_\gamma}
\eea
\noindent with $X=D$.
As found below, we predict $R$ to be about 5 times 
larger than the result of \cite{Aliev}.
Let us note here that the suggestion to use the ratio between
the Dalitz decay and the real
photon emission--and in particular its $q^2 \equiv m_{12}^2$
dependence--to determine
the quantum numbers of the
decaying particle, was made many years ago \cite{Feinberg,Landsberg}.
Our results are consistent with theirs.
\end{enumerate}

Since we will be mainly concerned with ratios $R$,
where $X=D,\,B$,
it is sufficient to parametrize the $X^*\to X\gamma$ transition with 
some effective coupling constant $g_{X^*X\gamma}$ and a 
suitable form factor ${\cal F}$.
We replace the various form factors possible 
for off--shell photons by a single one, and furthermore assume that it
is independent of $q^2$.
This is justified by the 
observation that for all the $X^*$--$X$ combinations we consider,
the mass difference $\Delta_{X^*X} \equiv m_{X^*}-m_X$ is 
of the order of up to $150$ MeV and thus
much smaller than the $\rho$--mass which is the most
relevant one in the vector dominance model for the form factors. 
Therefore the influence of the form factors on the results 
is indeed very small. We have confirmed that our numerical
results are practically unaffected by the 
assumption ${\cal F}(q^2) \approx {\cal F}(0)$.

The matrix elements we consider read \cite{Pilk}
\bea
{\cal M}_{1^-0^-\gamma} &=& g_{X^*X\gamma}\cdot{\cal F}(q^2)\cdot 
                                                      \epsilon^{\alpha\beta\mu\nu}
                                                      \epsilon_\alpha(\gamma)\epsilon^*_{\beta}(X^*)
                                                     P_\mu(X^*)q_\nu(\gamma)\nnb\\
{\cal M}_{2^+0^-\gamma} &=& g_{X^*X\gamma}\cdot{\cal F}(q^2)\cdot 
                                                      \epsilon^{\alpha\beta\mu\nu}
                                                      \epsilon_\alpha(\gamma) P_{\beta}(X^*)
                                                      \epsilon^*_{\mu\rho}(X^*)
                                                     q_\nu(\gamma) q^\rho(\gamma)\,.
\eea
\noindent We have concentrated here on the $1^-$ and $2^+$ quantum numbers for the excited and
$0^-$ for the ground state, since in the $D$--system the quantum numbers of the ground
states 
are known to be 
$0^-$ and because the $D^*$ decays into a $D$ and both a pion or a
photon. Therefore, the $1^-$ and $2^+$ are the lowest lying quantum numbers allowed
for the $D^*$. In fact, for the $B$--system the situation is slightly different. Here the small mass 
difference of the $B^*$ and the $B$ of roughly 45 MeV does not provide enough phase space for
the $B^*$ to decay into a $B$ and a $\pi$, hence the $B^*$ could in principle be a $1^+$.
This would cause the relevant
strong coupling constant $g_{B^*B\pi}$ to vanish.
However, we consider this idea to be too far--out and will not discuss 
it here.

We use the completeness relations (for the $2^+$ see e.g: \cite{TaoHan})
\bea
\sum\,\epsilon^{*\mu}(p)\epsilon^\nu(p) &=& 
-\left(g^{\mu\nu}-\frac{p^\mu p^\nu}{p^2}\right)\nnb\\
\sum\,\epsilon^{*\mu\nu}(p)\epsilon^{\rho\sigma}(p) &=& 
\frac{1}{2}\left[
\left(g^{\mu\rho}-\frac{p^\mu p^\rho}{p^2}\right)
\left(g^{\sigma\nu}-\frac{p^\sigma p^\nu}{p^2}\right)\nnb\right.\\
&&\left.\; +\left(g^{\sigma\mu}-\frac{p^\sigma p^\mu}{p^2}\right)
   \left(g^{\nu\rho}-\frac{p^\nu p^\rho}{p^2}\right)\right.\nnb\\
&&\left.\; -\frac23\left(g^{\mu\nu}-\frac{p^\mu p^\nu}{p^2}\right)
   \left(g^{\rho\sigma}-\frac{p^\rho p^\sigma}{p^2}\right)
\right]\,.
\eea
\noindent After squaring, summing and averaging we obtain for the
the decay into a real photon
\bea
3\sum \overline{\left|{\cal M}_{1^-0^-\gamma}\right|^2} &=& 
2g_{X^*X\gamma}^2\cdot{\cal F}^2(0)\cdot (Pq)^2\nnb\\
5\sum \overline{\left|{\cal M}_{2^+0^-\gamma}\right|^2} &=& 
g_{X^*X\gamma}^2\cdot{\cal F}^2(0)\cdot \frac{(Pq)^4}{P^2}\,,
\eea
\noindent where $P$ and $q$ are the momenta of the 
excited meson and the photon, respectively.
The resulting branching ratios are given by
\bea
\Gamma_{1^-\to 0^-\gamma} &=& g_{X^*X\gamma}^2\cdot{\cal F}^2(0)\cdot
\frac{(m_{X^*}^2-m_X^2)^3}{96\pi m_{X^*}^3}\nnb\\
\Gamma_{2^+\to 0^-\gamma} &=& g_{X^*X\gamma}^2\cdot{\cal F}^2(0)\cdot
\frac{(m_{X^*}^2-m_X^2)^5}{1280\pi m_{X^*}^5}\,.
\eea
\noindent With suitable replacements we recover the known result 
for the width of 
the decay $a_2\to \pi\gamma$ \cite{a2topigamma}.

For the decay into $e^+ e^-$ the polarization vector $\epsilon_\mu$ of the photon has
to be replaced by the lepton--current ~$e\bar u(e^-)\gamma_\mu u(e^+)$. 
Squaring the matrix elements and summing and averaging over polarizations yields
\bea
\lefteqn{3\sum\overline{\left|{\cal M}_{1^-0^-ee}\right|^2} = 
2g_{X^*X\gamma}^2\cdot\frac{{\cal F}^2(q^2)}{q^4}\cdot}\nnb\\
&& \left[4m_e^2\left((Pq)^2-P^2q^2\right)+
              q^2\left(2(Pp_e)^2+2(Pp_{\bar e})^2-P^2q^2\right)\right]
\eea
\bea
\lefteqn{5\sum\overline{\left|{\cal M}_{2^+0^-ee}\right|^2} =
g_{X^*X\gamma}^2\cdot\frac{{\cal F}^2(q^2)}{q^4}\cdot\frac{(Pq)^2-P^2q^2}{P^2}\cdot}\nnb\\
&&\left[4m_e^2\left((Pq)^2-P^2q^2\right)+
              q^2\left(2(Pp_e)^2+2(Pp_{\bar e})^2-P^2q^2\right)\right]\,.
\eea
\noindent The resulting widths  $\Gamma(X^* \to X e^+ e^-)$ and their 
respective ratios to the real photon widths
agree with the results given in \cite{Landsberg}. 
Our results for
the ratio $R$, defined in Eq. \ref{ratio}, 
in some meson systems are displayed in Table \ref{Results}.
All of them are of the order of $5\cdot 10^{-3}$.
\begin{table}[h]
\begin{center}
\begin{tabular}{|c||c|c|c|c|}
\hline
$\indvert X$ & 
$R$ ($1^-$) &
$R$ ($2^+$) &
${\cal B}(X^*\to X\gamma)$ & $ m_{X^*}-m_X$ \\[2mm]
\hline
$\indvert  B^0_s$         & 4.65 $\times 10^{-3}$& 4.34 $\times 10^{-3}$& 
                              dominant                   & $45.78 \pm 0.35$ MeV  \\     
$\indvert  B^0$             & 4.69 $\times 10^{-3}$& 4.38 $\times 10^{-3}$& 
                              dominant                   & $45.78\pm 0.35$ MeV  \\     
$\indvert  D_s^\pm$     & 6.45 $\times 10^{-3}$& 6.14 $\times 10^{-3}$& 
                              $0.942\pm 0.025$       & $143.8\pm 0.4$ MeV \\
$\indvert  D^0$             & 6.44 $\times 10^{-3}$& 6.13 $\times 10^{-3}$& 
                              $0.381\pm 0.029$       & $142.12\pm 0.07$ MeV \\
$\indvert  D^\pm$         & 6.42 $\times 10^{-3}$& 6.11 $\times 10^{-3}$& $0.011 {\begin{array}{l}
                                                                                 \scriptstyle +0.021 \\ \scriptstyle -0.007
                                                                                 \end{array}}$         & $140.64\pm 0.10$ MeV \\
$\indvert  K^0$            & 7.99 $\times 10^{-3}$& (7.68 $\times 10^{-3}$) & 
                              $0.0023\pm 0.0002$   & $398.42\pm 0.28$ MeV \\
\hline
\end{tabular}
\parbox{12cm}{\caption{\label{Results} \footnotesize
Predicted
ratios 
$R=\Gamma(X^* \to Xe^+e^-)/\Gamma(X^*\to X\gamma)$, 
for some mesonic systems $X$.
The branching ratios ${\cal B}(X^*\to X\gamma)$ 
and mass differences $m_{X^*}-m_X$ are from \cite{RPP}.
We have assumed $m_{B_s^*}-m_{B_s}=m_{B_d^*}-m_{B_d}$.
The number for the $K^*(892)$ decay is included for completeness only, since here form factors
might become important, and of course, the $K^*(892)$ is a $1^-$--particle. 
Note however that $K^*(892)\to K e^+ e^-$ has not been observed. The only
ratio for a heavy system $X$ observed so far is $B^*$ \cite{Eigen},
where $R=(4.7 \pm 1.1 \pm 0.9) \times 10^{-3}$.}} 
\end{center}
\end{table}
As can be observed from the table, the ratio $R$ 
may serve as an indicator for the $J^P$ quantum numbers of $X^*$ mesons,
which are believed to have $J^P=1^{-}$. At present, the only ratio
for which an experimental number
exists is  
$R=(4.7 \pm 1.1 \pm 0.9) \times 10^{-3}$, for $B^*$ decay \cite{Eigen}.
Although the central value agrees well with 
the quark model quantum numbers $J^P(B^*)=1^{-}$, it is premature to claim,
in view of the large error, a clear--cut rejection of the $2^+$ possibility.

\begin{figure}
\begin{center}
\includegraphics[height=12cm,angle=270]{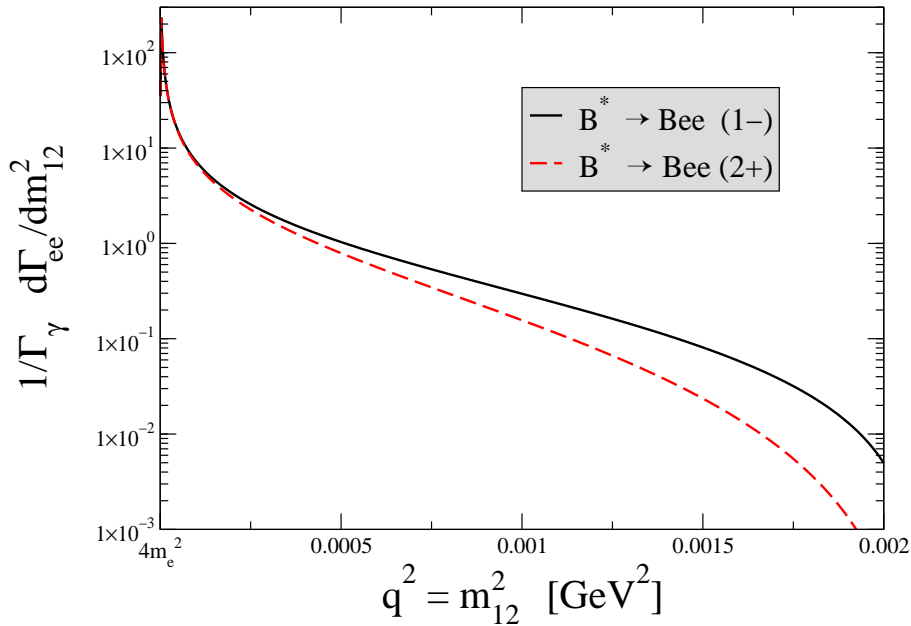}\\
\parbox{12cm}{\caption{\label{fig1} \footnotesize $m_{12}^2$--distribution 
of the electron--positron pairs in the
Dalitz decay $B^*\to Be^-e^+$ with $q$ the invariant mass of the pairs. Clearly, the quantum 
numbers affect the tail of the distribution.}}
\end{center}
\end{figure}

A better indicator for the quantum numbers of $X^*$ is the
distribution of the invariant mass squared ($q^2=m_{12}^2$) of the
$e^+ e^-$ pair.
To illustrate the effect of different quantum numbers on 
$d\Gamma(X^{*0} \to X^0 e^+ e^-)/ dq^2$, 
scaled by $\Gamma(B^{*0} \to B^0 \gamma)$,
we display in Fig.\,\ref{fig1} the results for
the $B$--system. Similar results are obtained for $X=D$.
Clearly, the change in the quantum numbers affects
the tail of the $q^2$--distribution of the electron--positron pairs 
by factors about 2 and larger. Modifications of
$q^2$ distribution depicted in Fig.\,\ref{fig1} by Vector Meson
Dominance
form factors would be practically invisible.
This provides an excellent tool to determine the quantum numbers of $X$.

To summarize, we have presented results for the decay widths of
$D^*\to De^+e^-$, $D_s^{*\pm}\to D_s^\pm e^+e^-$ and $B^{*0}\to B^0e^+e^-$.
All of them are of the order of roughly $0.5\%$ 
of the corresponding decays into real photons, as expected from a QED--like
calculation. The difference between the two $J^P$ assignments $1^-$ 
and $2^+$ for the the decaying particle, is about $5\%$. A significant
difference arises for the $q^2$--distribution, especially at high $q^2$.
In the light of the recent experimental results by \cite{Eigen} on the invariant
mass spectrum in the Dalitz decay of $B^*$'s we are very optimistic that the 
ambitious measurements we advertise are feasible.

\centerline{\bf Acknowledgments}
We thank H. Landsman, Y. Rozen and P. Singer for helpful discussions. 
The research of G.E. was supported in part by the Israel
Science Foundation founded by the Israel Academy of Sciences and
Humanities and by the Harry Werksman Research Fund.
F.K. would like to acknowledge financial 
support by the Minerva--foundation.

\end{document}